\documentclass[a4paper]{article}
\usepackage{amsmath,graphicx}

\usepackage{multirow}
\usepackage{booktabs}
\usepackage{array}
\usepackage{verbatim}
\newcolumntype{C}[1]{>{\centering\arraybackslash}p{#1}}
\RequirePackage[textfont=it,tableposition=top]{caption}

\usepackage{INTERSPEECH2019}

\title{The Microsoft System for VoxCeleb Speaker Recognition Challenge 2022}
\name{Gang Liu, Tianyan Zhou, Yong Zhao, Yu Wu, Zhuo Chen, Yao Qian, Jian Wu}
\address{
Microsoft Corporation, One Microsoft Way, Redmond, WA 98052, USA}
\email{\{liu.gang, tizhou, yonzhao, yuwu1, zhuc, yaoqian, jianwu\}@microsoft.com}

\begin{document}

\maketitle
\begin{abstract}
In this report, we describe our submitted system for track 2 of the VoxCeleb Speaker Recognition Challenge 2022 (VoxSRC-22).
We fuse a variety of good-performing models ranging from supervised models to self-supervised learning(SSL) pre-trained models. The supervised models, trained using VoxCeleb-2 dev data, consist of ECAPA-TDNN and Res2Net in a very deep structure. The SSL pre-trained models, wav2vec and wavLM, are trained using large scale unlabeled speech data up to million hours. These models are cascaded with ECAPA-TDNN and further fine-tuned in a supervised fashion to extract the speaker representations. All 13  models are applied with score normalization and calibration and then fused into the the submitted system. We also explore the audio quality measures in the calibration stage such as duration, SNR, T60, and MOS. The best submitted system achieves 0.073 in minDCF and 1.436\% in EER on the VoxSRC-22 evaluation set.
% Group 1 offers the more competitive individual system at the cost of intensive model complexity, while group 2 provides inferior metric at much less model complexity.
\end{abstract}

% Keywords command
\providecommand{\keywords}[1]
{
\textbf{{Index Terms:}} #1
}
%TC:ignore
%TC:endignore

%background:
\section{Introduction}
Recently, self-supervised learning (SSL) pre-training based models have demonstrated competitive performance in speaker recognition\cite{chen2022does,chen2022large, chen2022wavlm}. Canonical SSL models are trained either by discriminating the positive samples from the negative ones (e.g. wav2vec 2.0) [1], or by predicting discrete pseudo-labels on the masked regions (e.g. HuBERT) \cite{hsu2021hubert} and denoising modeling (e.g. wavLM) \cite{chen2022wavlm}. Both approaches leverage large amount of unlabeled speech data to implicitly capture phonetic information, which inevitably coupled with personal speaking style. At the same time, various residual blocks \cite{he2016deep} based network structures such as Res2Net \cite{zhou2021resnext} and ECAPA-TDNN \cite{desplanques2020ecapa}, can offer competitive performance at a much less computational cost than the SSL models. The submitted ensemble system leverages complementary information from multiple supervised and SSL models to tackle the challenge of the open condition competition.

\section{Proposed system}

Our submitted systems including three major components: supervised systems, SSL-based systems, and score calibration and fusion. In the SSL-based models, the upper stream SSL pre-trained models are trained with various unlabeled data. All the supervised systems (including the down-stream ECAPA-TDNN in SSL-based system) are trained with fix data set specified by the organizer. 

\subsection{Datasets}
\subsubsection{Training data and data augmentation}
The VoxCeleb2 dev dataset are used for all the supervised model training. It consists of 5,994 speakers and 1,092,009 utterances. Speed perturbation based augmentation is implemented by speed up and down 10\% via SoX tool. This gives 2x extra copy of speakers and utterances. The total speaker number in model training is 17,982 with 3,276,027 utterances. 
Kaldi-style data simulation \cite{povey2011kaldi} is implemented with four types of disturbance, babble, music, noise, and reverberation, which offers 4x of original data. All the above-mentioned data are offline data augmentation and simulation. Overall, the offline data size is 15x of original VoxCeleb2 dev dataset.

To further increase training data variations and model robustness, online simulation is also implemented and added at a given probability \cite{xiao2021microsoft}.

\subsubsection{Validation data}
We employ the following development sets to validate the proposed systems and  examine the settings of hyper parameter to avoid over-fitting. 
\begin{itemize}
   % \item Vox1-O: A small but well-cited validation set based on VoxCeleb1 test dataset with 40 speakers.
   % \item Vox2-test: A small validation set from VoxCeleb2 test dataset with self composed trials based on 120 speakers.
   \item VoxSRC21-val: Validation set of track 1 \& 2 of VoxSRC-21 with a focus on the language impact.
   \item VoxSRC22-val: Validation set of track 1 \& 2 of VoxSRC-22 with a focus on the age and noise impact.
 \end{itemize}

%Their statistics are summarized in Table~\ref{tab:validationSets}. 
% It is noted there is no speaker overlap between these various validation data sets and model training data.

\begin{comment}
\begin{table}[tb]
	\caption{Statistics of the validation set of track2 of VoxSRC-22 and VoxSRC-21. (*) Note, the speaker count for VoxSRC22-val dose not include 6,683 VoxConverse utterance.}
	\label{tab:validationSets}
	\centering
	\begin{tabular}{|c|c|c|c|c|c|}
		\hline
		\textbf{dataset} & \textbf{\#utt} & \textbf{\#trial}& \textbf{\#target} & \textbf{\#non} & \textbf{\#spk} \\
				\hline
				vox1-O & 4708 & 37611 & 18802 & 18809 & 40 \\ \hline
Vox2-test & 26591 & 30000 & 15190 & 14810 & 120 \\ \hline
VoxSRC21-val & 64711 & 60000 & 29969 & 30031 & 1251 \\ \hline
VoxSRC22-val & 110366 & 306432 & 159789 & 146643 & *1205 \\
		\bottomrule
	\end{tabular}
\end{table}
\end{comment}

\begin{table*}[tb]
	\caption{Evaluation results on VoxSRC21-Val and VoxSRC22-val test set. wavLM and wav2vec are SSL-based systems, the rest are supervised systems. The training data for system 1 are 94kh public data, which are augmented with noise and used as training data for system 2. System 3-5 differ in learning rate (lr) and epoch setting in large margin finetune stage, which are trained with unlabeled data significantly larger than previous published wavLM. System 6-7 using same models but with different cohort size in scoring stage, the SSL-model is trained with 56kh data. System 9-10 differs in training finetune steps}
	\label{tab:summary-result}
	\centering
	\begin{tabular}{|c|c|c|c|c|c|}
		\hline
\multirow{2}{*}{\textbf{No.}} &
\multirow{2}{*}{\textbf{System}} &
		\multicolumn{2}{c|}{\textbf{VoxSRC22-val}} &
		\multicolumn{2}{c|}{\textbf{VoxSRC21-val}} \\ 
		\cline{3-6}
		& & \textbf{EER(\%)} & \textbf{DCF} &  \textbf{EER(\%)} & \textbf{DCF} \\
				\hline
1 & wavLM-large-v1 & 1.44 & 0.0944 & 1.92 & 0.1205  \\
2 & wavLM-large-v2 & 1.39 & 0.0902 & 1.88 & 0.1116  \\
3 & wavLM-large-v3 & 0.95 & 0.0658 & 1.31 & 0.0721  \\
4 & wavLM-large-v4 & 1.11 & 0.0744 & 1.52 & 0.0879  \\
5 & wavLM-large-v5 & 0.92 & 0.0630 & 1.37 & 0.0797  \\
6 & wav2vec2.0-XLSR2-v1 & 1.25 & 0.0816 & 1.67 & 0.1027  \\
7 & wav2vec2.0-XLSR2-v2 & 1.25 & 0.0847 & 1.59 & 0.0971  \\
\hline
8 & Res2Net-50 & 1.36 & 0.0863 & 1.56 & 0.0891  \\
9 & Res2Net-101-v1 & 1.24 & 0.0857 & 1.65 & 0.0984  \\
10 & Res2Net-101-v2 & 1.23 & 0.0834 & 1.63 & 0.0986  \\
11 & Res2Net-152 & 1.25 & 0.0866 & 1.74 & 0.0977  \\
12 & ECAPA-TDNN(C512) & 1.88 & 0.1305 & 2.54 & 0.1509  \\
13 & ECAPA-TDNN(C1024) & 1.65 & 0.1039 & 2.15 & 0.1302  \\
\hline
14 & fusion of 13 models & 0.81 & 0.0518 & 1.12 & 0.0585  \\
15 & fusion of 13 models + 3 audio quality measures & 0.79 & 0.0513 & 1.10 & 0.0565  \\
		\bottomrule
	\end{tabular}
\end{table*}

\subsection{Supervised systems}
Our supervised systems consist of Res2Net and ECAPA-TDNN. The ResNet-based architecture \cite{he2016deep} has been widely adopted
to extract speaker embeddings for text-independent speaker
veriﬁcation systems. The Res2Net \cite{zhou2021resnext} redesigns the residual block. It constructs hierarchical residual-like connections inside the residual block and assembles variable-size receptive ﬁelds within one layer. With the scale dimension, the Res2Net model can represent multi-scale features with various granularity, which facilitates speaker veriﬁcation for very short utterances.
In this challenge, we include three Res2Net networks with different depths, i.e. Res2Net-50, Res2Net-101 and Res2Net-152. 
All convolutional layers are followed by a batch normalization layer and rectified linear units (ReLU) activation function. The models take 80-dimensional log Mel ﬁlter banks as input features and generate speaker embedding with a dimension of 256. 
%For ResNet structure, we adopt the standard ResNet-152 with a starting width of 64.
For ECAPA-TDNN, we implement both smaller (C=512) and larger (C=1024) versions to provide complementary decisions to the Res2Net structures during system fusion stage.  

All the supervised models are trained with the large margin strategy \cite{thienpondt2021} in two stages as following:
\begin{itemize}
   \item \textbf{Stage 1}: In this stage, we use the 3x speed perturbed data to augment training, which means the total number of speakers is 17,982. The training segments are randomly sampled in 2s chunks. The margins used for AM loss and AAM loss are 0.2 and 0.3 respectively. We train this stage for 150 epochs.
   \item \textbf{Stage 2}: in this stage, we remove the speed perturbed data and keep the original 5,994 speakers. The training segments are increased to 6s chunks. The margins used for AM loss and AAM loss are also increased to 0.4 and 0.5 respectively. We train this stage for 3 epochs.
 \end{itemize}

\begin{table}[tb]
	\caption{ Ablation Study on Single System (i.e. system 5 in Table 1). '+' denotes stacking on previous method(s).}
	\label{tab:singleSystemOptimization}
	\centering
	\begin{tabular}{|l|c|c|c|c|}
		\hline
\multirow{2}{2em}{\textbf{scoring}} &
		\multicolumn{2}{c|}{\textbf{VoxSRC21-val}} &
		\multicolumn{2}{c|}{\textbf{VoxSRC22-val}} \\
		\cline{2-5}
		& \textbf{EER(\%)} & \textbf{DCF} & \textbf{EER(\%)} & \textbf{DCF} \\
\hline
Cosine & 1.71 & 0.099 & 1.20 & 0.083 \\
+ S-Norm & 1.53 & 0.084 & 1.09 & 0.073 \\
++ Calib & 1.44 & 0.080 & 0.92 & 0.063 \\
		\bottomrule
	\end{tabular}
\end{table}

\subsection{SSL-based systems}
In SSL-based systems, the input is raw audio, the output from SSL pre-trained model is low dimension embedding which is then used as feature and fed into downstream supervised system for speaker verification system. In this study, two types of SSL-based models are explored, i.e., wavLM and wav2vec 2.0, see the first part in Table~\ref{tab:summary-result}. Five wavLM-based models \cite{chen2022wavlm} are produced. They are in the same model structure, but differ in training data sources and training settings.  
The model wavLM-large-v1 is pre-trained with 94k hours of English data, including LibriLight,VoxPopuli, and GigaSpeech \cite{chen2022wavlm}. The model wavLM-large-v2 is trained with the same data as wavLM-large-v1, but augmented with noisy speech simulation. The models wavLM-large-v3/v4/v5 are trained with a large scale in-house dataset with hundreds of thousands hours of speech audio. 
The three models stem from the same wavLM model, but differ in the learning rate (lr) and epoch settings in large margin fine-tune stage.

Moreover we trained two wav2vec-based models using multilingual speech data set with more than 36 languages. The pre-trained model Wav2vec2.0 Large (XLSR) is downloaded via Fairseq sequence modeling toolkit \cite{ott2019fairseq}.
All the models are trained with Additive Angular Margin Loss (AAM) \cite{deng2019arcface}, in which the margin is set as 0.2 unless specified otherwise. Training segments are randomly sampled 3s chunks from each audio utterance unless specified otherwise.
The concatenated system of upper stream SSL model and down stream ECAPA-TDNN (C=512) are finetuned with three stages as following:
\begin{itemize}
   \item Stage 1: the concatenated system are trained for 10 epochs during which the upper stream SSL model parameter are fixed.
   \item Stage 2: the concatenated system are trained for 5 epochs during which both upper and down stream are allow to be updated.
   \item Stage 3: the concatenated system are trained for 2 to 3 epochs during which both upper and down stream are allow to be updated. Training segments are randomly sampled 6s chunks from each audio utterance, the margin in AAM loss is set as 0.4-0.5.
 \end{itemize}

Once the model training is finished, the 256-dimension embeddings extracted from the model are used to perform score normalization and calibration, which can improve system performance step-by-step, which is detailed in Table~\ref{tab:singleSystemOptimization}

\begin{comment}
\begin{table*}[tb]
	\caption{ Ablation Study on Single System (system 3 in Table 3). '+' denotes stacking on previous method(s).}
	\label{tab:singleSystemOptimization}
	\centering
	\begin{tabular}{|c|c|c|c|c|c|c|c|c|}
		\hline
\multirow{2}{2em}{\textbf{scoring}} &
		\multicolumn{2}{c|}{\textbf{Vox1-O}} &
		\multicolumn{2}{c|}{\textbf{Vox2-test}} &
		\multicolumn{2}{c|}{\textbf{VoxSRC21-val}} &
		\multicolumn{2}{c|}{\textbf{VoxSRC22-val}} \\
		\cline{2-9}
		& \textbf{EER(\%)} & \textbf{DCF} & \textbf{EER(\%)} & \textbf{DCF} & \textbf{EER(\%)} & \textbf{DCF} & \textbf{EER(\%)} & \textbf{DCF} \\
\hline
CDS & 0.43 & 0.028 & 2.12 & 0.055 & 1.71 & 0.099 & 1.20 & 0.083 \\
\hline
+SNORM & 0.38 & 0.024 & 2.11 & 0.052 & 1.53 & 0.084 & 1.09 & 0.073 \\
\hline
+Calibration & 0.28 & 0.021 & 2.10 & 0.048 & 1.44 & 0.080 & 0.92 & 0.063 \\
		\bottomrule
	\end{tabular}
\end{table*}
\end{comment}

\subsection{Score calibration and fusion}
Scores from each individual system go through score normalization and calibration \cite{thienpondt2021idlab}. The ensemble system fuse scoring results from multiple systems. The core parts can be highlighted as following:
\begin{itemize}
   \item S-Norm: the utterance from the official VoxCeleb2 dev dataset are chosen for each speaker, their embedding are averaged to derived 5,994 cohorts. The cosine distance score will be normalized by top N impost scores. N is chosen as 200-600 in this study.

   \item Score calibration: Logistic regression model are adopted for calibration, VoxSRC-22 is used as development set. The score after S-Norm, duration, embedding magnitude and their function (such as min/max value, difference) are used as feature for training and testing.

   \item Fusion: During ensemble system fusion stage, a logistic regression model is trained on VoxSRC-22 with all scores from various sub-system, which is then applied to scores on VoxSRC-22 blind data set to fuse different system. The hyper parameter tuning are performed on the validation set of VoxSRC-22 and VoxSRC-21.
   To further boost speaker verification performance, a extra series of quality measures \cite{thienpondt2021idlab}\cite{garcia2006using} are explored during score calibration and fusion stage, which consist of T60, Mean opinion score (MOS) and their functions. 
 \end{itemize}

\begin{table}[tb]
	\caption{Evaluation results of two submissions on VoxSRC22-Val and VoxSRC22-blind sets. }
	\label{tab:summary-result2}
	\centering
	\setlength\tabcolsep{3pt} % default value: 6pt
	\begin{tabular}{|c|c|c|c|c|c|}
		\hline
		\multirow{2}{*}{\textbf{System}} &
		\multicolumn{2}{c|}{\textbf{VoxSRC22-val}} &
		\multicolumn{2}{c|}{\textbf{VoxSRC22-blind}} \\ 
		\cline{2-5}
		& \textbf{EER(\%)} & \textbf{DCF} &  \textbf{EER(\%)} & \textbf{DCF} \\
				\hline
				wavLM-large-v3 & 0.950 & 0.0658 & 1.606 & 0.0921 \\
                Fusion of 13 models & 0.790 & 0.0513 & 1.436	& 0.0728 \\
		\bottomrule
	\end{tabular}
\end{table}

\section{Experimental results}
\label{sec:experiments}
We evaluated the proposed approach on the VoxSRC21-val and VoxSRC22-val sets. Table~\ref{tab:summary-result} shows the performance of individual models and the fused system. We take the minDCF of VoxSRC22-val as primary performance indicator. 
It shows that model \#5 (wavLM-large-v5) produces the best performance of 0.630 in minDCF among all individual models. The models wavLM-large-v3/v4/v5 outperform the wavLM-large-v1/v2 due to the use of larger scale pre-training data. The wav2vec models yields minDCF of 0.0816 (model \#6), worse than the best-performing wavLM model. 
The SSL models, especially wavLM-large-v3/v4/v5, outperforms the supervised models by a significant margin. The Res2Net-101-v2 produces the best minDCF of 0.0834 among all the supervised models. The Re2Net models performs better than the ECAPA-TDNN models.

The ensemble of all the 13 models produces 0.0518 in minDCF on the VoxSRC22-val set, 18\% relative improvement over the best individual system (model \#5). 
By enclosing the audio quality features in the fusion model, The minDCF is further improved to 0.0513. 

Table~\ref{tab:summary-result2} shows the results of two submissions in the Challenge. The best fusion system achieves 0.073 in minDCF and 1.436\% in EER on the VoxSRC22-blind set, which reduces minDCF by 18.9\% relative compared with the single model (wavLM-large-v3).

\section{Conclusions}
In this challenge, we exploit a variety of supervised and SSL-based models. 
The SSL-based models take advantage of the speech representation capabilities learned from large amount of unlabeled data, yielding superior performance over the conventional supervised models. 
We applied score level fusion to leverages complementary information of these models. We also explored the audio quality measures in the calibration stage such as duration, SNR, T60, and MOS. The best submitted system achieves 0.073 in minDCF and 1.436\% in EER on the VoxSRC-22 evaluation set.

\bibliographystyle{IEEEtran}

\bibliography{mybib}

% \begin{thebibliography}{9}
% \bibitem[1]{Davis80-COP}
%   S.\ B.\ Davis and P.\ Mermelstein,
%   ``Comparison of parametric representation for monosyllabic word recognition in continuously spoken sentences,''
%   \textit{IEEE Transactions on Acoustics, Speech and Signal Processing}, vol.~28, no.~4, pp.~357--366, 1980.
% \bibitem[2]{Rabiner89-ATO}
%   L.\ R.\ Rabiner,
%   ``A tutorial on hidden Markov models and selected applications in speech recognition,''
%   \textit{Proceedings of the IEEE}, vol.~77, no.~2, pp.~257-286, 1989.
% \bibitem[3]{Hastie09-TEO}
%   T.\ Hastie, R.\ Tibshirani, and J.\ Friedman,
%   \textit{The Elements of Statistical Learning -- Data Mining, Inference, and Prediction}.
%   New York: Springer, 2009.
% \bibitem[4]{YourName17-XXX}
%   F.\ Lastname1, F.\ Lastname2, and F.\ Lastname3,
%   ``Title of your INTERSPEECH 2019 publication,''
%   in \textit{Interspeech 2019 -- 20\textsuperscript{th} Annual Conference of the International Speech Communication Association, September 15-19, Graz, Austria, Proceedings, Proceedings}, 2019, pp.~100--104.
% \end{thebibliography}

\end{document}